\documentclass[ACS,STIX1COL]{WileyNJD-v2}

\usepackage{color}

\usepackage{mathtools}
\usepackage{graphicx}
\usepackage{dcolumn}
\usepackage{array}
\usepackage{lipsum}
\usepackage{bm}
\usepackage{subfigure}
\usepackage{multirow}
\usepackage{tabularx}
\usepackage{amsmath}
\usepackage{braket}
\graphicspath{{plots/}}

\newcommand{\beq}{\begin{equation}}
\newcommand{\eeq}{\end{equation}}
\newcommand{\bea}{\begin{eqnarray}}
\newcommand{\eea}{\end{eqnarray}}

\articletype{ORIGINAL ARTICLE}%

\received{xxx}
\revised{xxx}
\accepted{xxx}

\begin{document}

\title{Nonlinear electronic density response of the ferromagnetic uniform electron gas at warm dense matter conditions}

\author[1,2]{Tobias Dornheim$^\dagger$}

\author[1,2]{Zhandos A. Moldabekov}

 \author[2]{Jan Vorberger}

\authormark{Tobias Dornheim \textsc{et al}}

\address[1]{\orgdiv{Center for Advanced Systems Understanding (CASUS)}, \orgname{} \orgaddress{\state{D-02826 G\"orlitz}, \country{Germany}}}

\address[2]{\orgdiv{Helmholtz-Zentrum Dresden-Rossendorf (HZDR)}, \orgname{} \orgaddress{\state{D-01328 Dresden, Germany}, \country{Germany}}}


 \corres{$^\dagger$\email{t.dornheim@hzdr.de}}


\abstract{
In a recent Letter [T.~Dornheim \emph{et al.}, Phys.~Rev.~Lett.~\textbf{125}, 085001 (2020)], we have presented the first \emph{ab initio} results for the nonlinear density response of electrons in the warm dense matter regime. In the present work, we extend these efforts by carrying out extensive new path integral Monte Carlo (PIMC) simulations of a \emph{ferromagnetic} electron gas that is subject to an external harmonic perturbation. This allows us to unambiguously quantify the impact of spin-effects on the nonlinear density response of the warm dense electron gas. In addition to their utility for the description of warm dense matter in an external magnetic field, our results further advance our current understanding of the uniform electron gas as a fundamental model system, which is important in its own right.
}

\keywords{warm dense matter, nonlinear density response, path integral Monte Carlo\\\vspace*{0.2cm}\textbf{This work is dedicated to the late Vladimir Fortov.}}

\maketitle



\section{Introduction\label{sec:intro} }

The study of matter under extreme conditions~\cite{fortov_review} has emerged as one of the most active frontiers in physics, material science, and related disciplines over the last decades. Of particular interest is so-called warm dense matter (WDM)---an exotic state that is characterized by high temperatures ($T\sim10^3-10^8K$) and densities exceeding that of solids~\cite{new_POP,wdm_book}. In nature, WDM conditions occur in astrophysical objects such as giant planet interiors~\cite{Militzer_2008,militzer1} and the crust of neutron stars~\cite{Daligault_2009}. Furthermore, WDM has been predicted to occur on the implosion path of a fuel capsule in inertial confinement fusion applications~\cite{hu_ICF}, and can be used to accelerate chemical reactions~\cite{Brongersma2015}. For these reasons, WDM is nowadays routinely realized in large research facilities around the world using different experimental techniques; see the recent review by Falk~\cite{falk_wdm} for an accessible overview.

From a theoretical perspective, WDM conditions are defined by two characteristic parameters that are both of the order of one at the same time: a) the density parameter (also known as Wigner-Seitz radius) $r_s=\overline{r}/a_\textnormal{B}$, where $\overline{r}$ and $a_\textnormal{B}$ are the mean inter-particle distance and (first) Bohr radius, and b) the degeneracy temperature $\theta=k_\textnormal{B}T/E_\textnormal{F}$, where $E_\textnormal{F}$ corresponds to the usual Fermi energy~\cite{Ott2018}. In particular, the highly nontrivial interplay of i) Coulomb coupling, ii) thermal excitations, and iii) quantum degeneracy effects such as Pauli blocking renders the accurate description of WDM a most formidable challenge~\cite{review,new_POP,wdm_book}. Consequently, no single numerical or analytical method is capable to describe all aspects of WDM over the relevant parameter range.

In addition, contemporary WDM theory is mostly based on linear response theory (LRT)~\cite{nolting,quantum_theory}, i.e., the assumption of a weakly perturbed system that reacts in first-order only to an external perturbation. Specific examples for the application of LRT include the interpretation of X-ray Thomson scattering experiments~\cite{siegfried_review,kraus_xrts,dornheim_PRL_ESA_2020}, the construction of exchange--correlation functionals for density functional theory~\cite{Karasiev_PRL_2018,pribram,Sjostrom_PRB_2014}, and the construction of effective electronically screened ionic potentials~\cite{zhandos1,zhandos2,Ceperley_Potential_1996}. On the one hand, the assumption of a linear response leads to a drastic simplification of the theoretical description, and, in this way, makes many problems tractable. In addition, many linear-response properties of electrons in the WDM regime have recently become available based on \emph{ab initio} path integral Monte Carlo (PIMC) simulations~\cite{dornheim_ML,dornheim_PRL_ESA_2020,Dornheim_PRB_ESA_2021,dornheim_dynamic,dynamic_folgepaper,Dornheim_PRE_2020,Hamann_PRB_2020,groth_jcp,dornheim_pre}. On the other hand, the validity of LRT has rarely been checked, which introduces a potential source of systematic errors.

Recently, Dornheim \textit{et al.}~\cite{Dornheim_PRL_2020} have presented the first data for the \emph{nonlinear density response} of WDM, by carrying out extensive PIMC simulations of a harmonically perturbed, inhomogeneous electron gas. First and foremost, this has allowed to unambiguously check the range of applicability of LRT with respect to the strength of the external perturbation. Indeed, it was found that nonlinear effects cannot be neglected in many situations of experimental relevance. Furthermore, they have obtained the first numerical results for the cubic density response of the UEG for different densities and temperatures. Subsequently, the same group has presented the backbone of a new theory of the nonlinear density response of WDM~\cite{Dornheim_PRR_2021}, which is based on the highly accurate representation of the static local field correction~\cite{dornheim_ML,dornheim_PRL_ESA_2020,Dornheim_PRB_ESA_2021} and which extends the previous results obtained using the mean-field approximation \cite{Rostami}. More specifically, this theory does not only cover the wave-number and frequency of the original perturbation, but also covers the excitation of higher harmonics. In fact, the quadratic density response at the second harmonic constitutes the dominant nonlinear contribution at intermediate perturbation amplitudes~\cite{Dornheim_PRR_2021}.

Despite all these recent advances, a study of the impact of the spin-polarization $\xi=(N^\uparrow-N^\downarrow)/N$, where $N$, $N^\uparrow$, and $N^\downarrow$ denote the total number of electrons, number of majority electrons, and number of minority electrons, respectively, has hitherto been missing. This is unfortunate, as spin-effects are known to play an important role for WDM in an external magnetic field~\cite{dornheim2021momentum,Appelbe,Haensel,Lifshitz}. Other examples include the construction of exchange--correlation functionals for spin density functional theory~\cite{ksdt,groth_prl,review,vwn}, and the description of atoms and molecules like oxygen.

In the present work, we partly remedy this shortcoming by carrying out extensive direct PIMC simulations of a harmonically perturbed \emph{ferromagnetic} electron gas, i.e., for $\xi=1$. This allows us to unambiguously characterize the impact of the spin-polarization on the nonlinear density response of electrons in the WDM regime, which i) cannot be neglected and ii) depends on the wave-number of the perturbation. In addition to their value for WDM theory, these results further advance our current understanding of the uniform electron gas as a fundamental model system~\cite{status,groth_prl,dornheim_prl,review,loos,quantum_theory}.

The paper is organized as follows: In Sec.~II, we introduce the relevant theoretical background of the PIMC estimation of the nonlinear density response. The subsequent Sec.~III contains our new simulation results and the corresponding analysis of the impact of spin-effects. The paper is concluded by a brief summary and outlook in Sec.~IV.

\section{Theory\label{sec:theory}}


Throughout this work, we employ the direct PIMC method~\cite{Takahashi_Imada_PIMC_1984,Berne_JCP_1982} as it has been described, e.g., in the review article by Ceperley~\cite{cep}. Specifically, we employ a canonical implementation~\cite{mezza} of the worm algorithm by Boninsegni \emph{et al.}~\cite{boninsegni1,boninsegni2}, which ensures an efficient sampling of the fermionic permutation space~\cite{Dornheim_JCP_2020}. In addition, we stress that we do not impose any nodal restrictions~\cite{Ceperley1991} in our approach. Therefore, these exact direct PIMC simulations are afflicted with the notorious fermion sign problem~\cite{dornheim_sign_problem,dornheim2021fermion}, which constitutes the main limitation in practice. 

Following the procedure introduced in Refs.~\cite{Dornheim_PRL_2020,bowen2,moroni,moroni2,groth_jcp,dornheim_pre}, we simulate a harmonically perturbed electron gas that is governed by the Hamiltonian
\begin{eqnarray}\label{eq:hamiltonian}
\hat H = \hat H_\textnormal{UEG} + 2 A \sum_{l=1}^N \textnormal{cos}\left( \hat{\mathbf{r}}_l\cdot{\mathbf{q}} \right)\ ,
\end{eqnarray}
and we use Hartree atomic units throughout this work.
Specifically, $\hat H_\textnormal{UEG}$ denotes the usual Hamiltonian of the uniform electron gas (see Ref.~\cite{review} for details), and the perturbation amplitude $A$ and wave vector $\mathbf{q}$ characterize the external perturbation.
The density response of the system can then be straightforwardly estimated by evaluating the expectation value of the density operator in reciprocal space at a wave vector $\mathbf{k}$,
\begin{eqnarray}\label{eq:rho}
\braket{\hat\rho_\mathbf{k}}_{q,A} = \frac{1}{V} \left< \sum_{l=1}^N e^{-i\mathbf{k}\cdot\hat{\mathbf{r}}_l} \right>_{q,A} \ , 
\end{eqnarray}
where $\braket{\dots}_{q,A}$ indicates that the expectation value is computed with respect to Eq.~(\ref{eq:hamiltonian}).
For the first two harmonics of the original perturbation, i.e., for $\mathbf{k}_1=\mathbf{q}$ and $\mathbf{k}_2=2\mathbf{q}$, Eq.~(\ref{eq:rho}) can be expanded as~\cite{Dornheim_PRR_2021}
\begin{eqnarray}\label{eq:rho1}
\braket{\hat\rho_\mathbf{q}}_{q,A} &=& \chi^{(1)}(q) A + \chi^{(1,\textnormal{cubic})}(q) A^3\ ,\\
\label{eq:rho2}
\braket{\hat\rho_\mathbf{2q}}_{q,A} &=& \chi^{(2)}(q) A^2 \ ,
\end{eqnarray}
where $\chi^{(1)}(q)$, $\chi^{(1,\textnormal{cubic})}(q)$, and $\chi^{(2)}(q)$ denote the linear response function, the cubic response function at the first harmonic, and the quadratic response function at the second harmonic, respectively.

In addition, it is useful to consider the density in coordinate space, $n(\mathbf{r})$, which can be expressed as a sum over all harmonics,
 \begin{eqnarray}\label{eq:expansion}
 n(\mathbf{r}) = n_0 + 2 \sum_{\eta=1}^\infty
 \braket{\hat\rho_{\eta\mathbf{q}}}_{q,A}
 \textnormal{cos}\left(
 \eta\mathbf{q}\cdot\mathbf{r}
 \right)\ .
 \end{eqnarray}

\section{Results\label{sec:results}}

\begin{figure}
    \centering
    \includegraphics[width=0.48\textwidth]{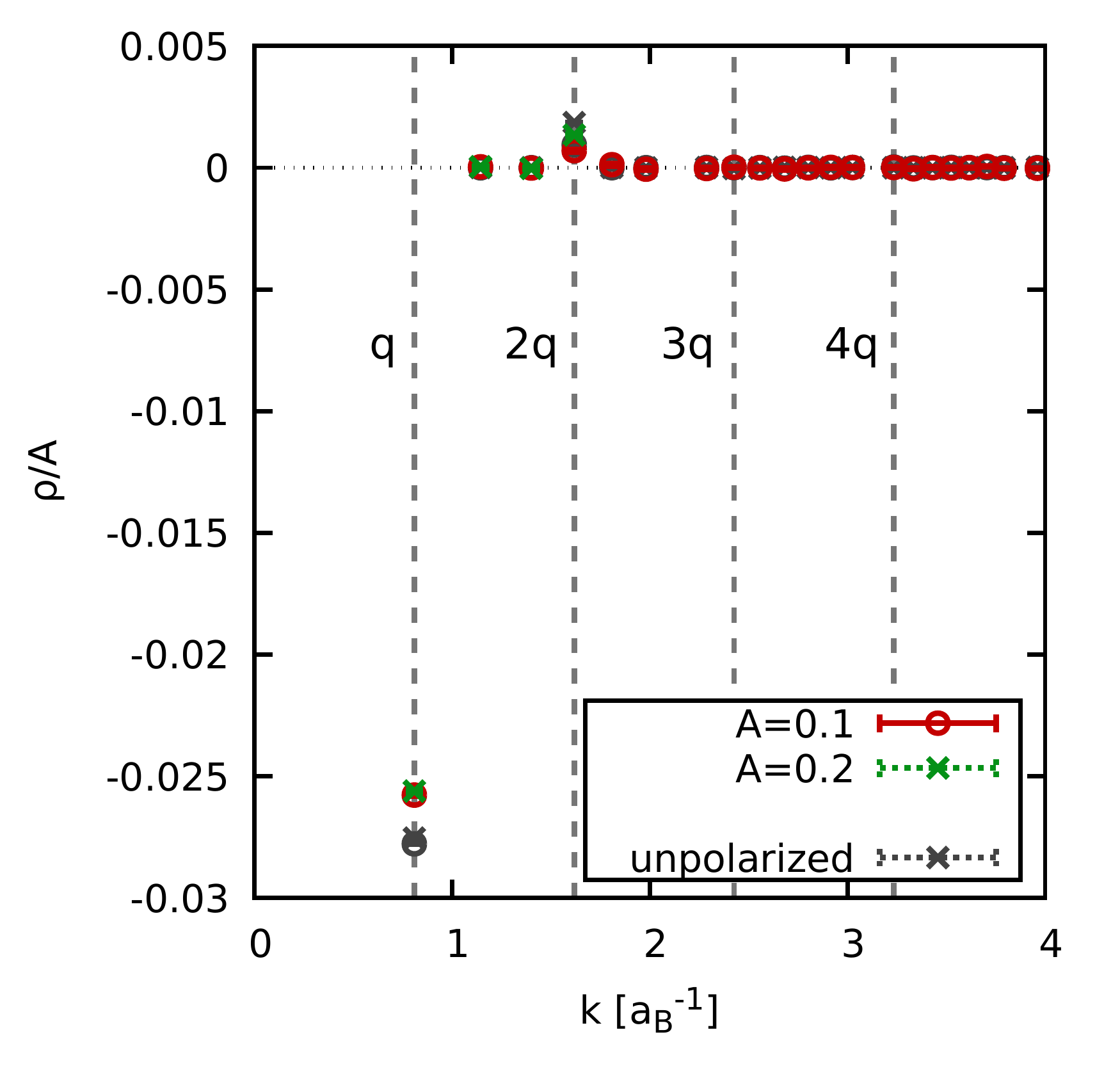}\includegraphics[width=0.48\textwidth]{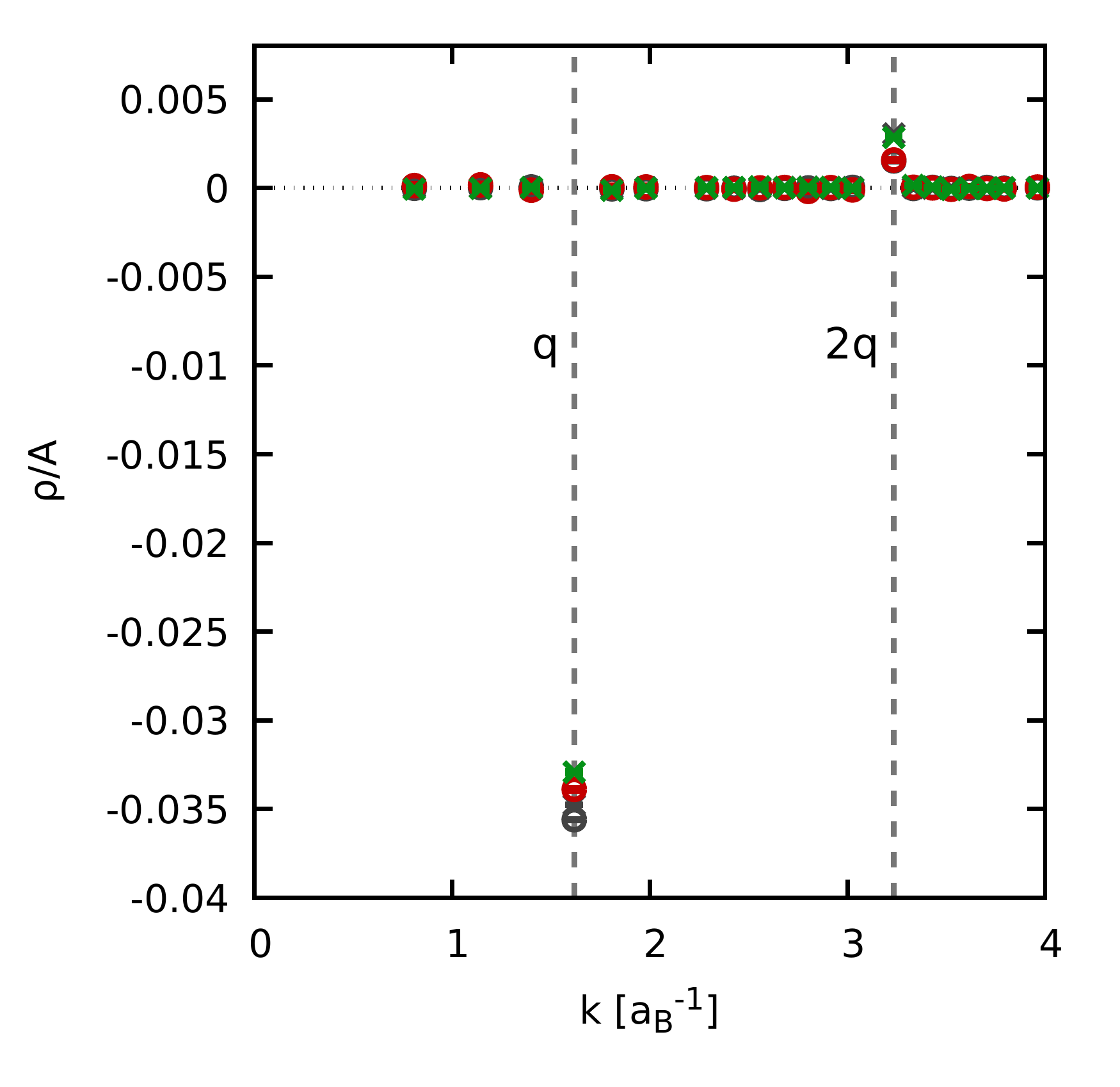}
    \caption{\label{fig:spectrum}Full wave-number dependence of the density response $\braket{\hat\rho_\mathbf{k}}_{q,A}$ of the UEG for $N=14$, $r_s=2$, and $\theta=1$ with $q=2\pi/L$ (left) and $q=4\pi/L$ (right). The coloured and grey symbols depict our new data for $\xi=1$ and data for $\xi=0$ taken from Ref.~\cite{Dornheim_PRR_2021}.
    }
\end{figure}

Let us begin our investigation of the nonlinear density response of the ferromagnetic electron gas at WDM conditions by considering the full spectrum of excitations in reciprocal space. This is shown in Fig.~\ref{fig:spectrum} for $N=14$ spin-polarized electrons at $r_s=2$ and $\theta=1$. These conditions are of prime importance for contemporary WDM research and can be realized, for example, in experiments with aluminum~\cite{Sperling_PRL_2015,Ramakrishna_PRB_2021}. Further, we note that we define $\theta$ with respect to the Fermi energy of the \emph{unpolarized} system, $E_\textnormal{F}=q_\textnormal{F}^2/2$, with
\begin{eqnarray}
q_\textnormal{F} = \left(
\frac{9\pi}{4}
\right)^{1/3} \frac{1}{r_s} \ .
\end{eqnarray}
This is advantageous as it allows us to compare our new data for the ferromagnetic case to previous results for the paramagnetic UEG at precisely the same conditions. On the other hand, this results in a substantially more severe manifestation of the fermion sign problem for the ferromagnetic case as all $N$ electrons can potentially exchange with each other. More specifically, we find an average sign of $S\sim10^{-2}$ ($S\sim10^{-1}$) for the ferromagnetic (paramagnetic) case, which means that the computational effort is increased by a factor of $100$ compared to previous investigations. A more detailed discussion of the fermion sign problem is beyond the scope of the present work, and the interested reader is referred to Ref.~\cite{dornheim_sign_problem}.

The left panel of Fig.~\ref{fig:spectrum} shows the spectrum for a perturbation of $\mathbf{q}=2\pi/L(1,0,0)^T$ ($q=|\mathbf{q}|=0.84q_\textnormal{F}$), i.e., for the smallest wave-vector that is possible in this particular finite cubic simulation cell of length $L$. The red circles and green crosses show our new PIMC results for $A=0.1$ and $A=0.2$, and the corresponding dark grey symbols depict the same information for the fully unpolarized case.  First and foremost, we note that all data sets exhibit a qualitatively similar behaviour, with a strong negative signal at the original perturbation and a substantially smaller signal at the second harmonic. For all other wave vectors $\mathbf{k}$, the response vanishes within the given Monte Carlo error bars.
In particular, no density response can be resolved for the third and fourth harmonics for these moderate values of $A$. In addition, we find that the signal is larger for the paramagnetic case at both the first and second harmonic, and for both depicted values of the perturbation amplitude $A$. This can be understood as follows: identical fermions are intrinsically correlated to each other by the Pauli exclusion principle, which prevents them from occupying the same position in coordinate space. Naturally, this effect is less pronounced for the paramagnetic case, as only half the particles are affected by this degeneracy pressure. As a consequence, the system is stiffer for $\xi=1$ compared to $\xi=0$, which, in turn, means that it reacts less strongly to the external perturbation.

The right panel of Fig.~\ref{fig:spectrum} shows the same information, but for the perturbation wave-vector $\mathbf{q}=2\pi/L(2,0,0)^T$ ($q=1.69q_\textnormal{F}$). Firstly, we find a similar behaviour as in the left panel, although the density response is more pronounced at both harmonics. In fact, it is well known that all three response functions introduced in Eq.~(\ref{eq:rho1}) exhibit a maximum around $q=2q_\textnormal{F}$, which is explained in detail in Ref.~\cite{Dornheim_PRR_2021}. Furthermore, we find that the deviations between the ferromagnetic and paramagnetic cases are less pronounced than for the smaller wave number. This nontrivial effect is discussed in more detail in the context of Figs.~\ref{fig:A_dependence_qx1} \& \ref{fig:A_dependence_qx2} shown below.

\begin{figure}
    \centering
   \hspace*{-0.03\textwidth} \includegraphics[width=0.51\textwidth]{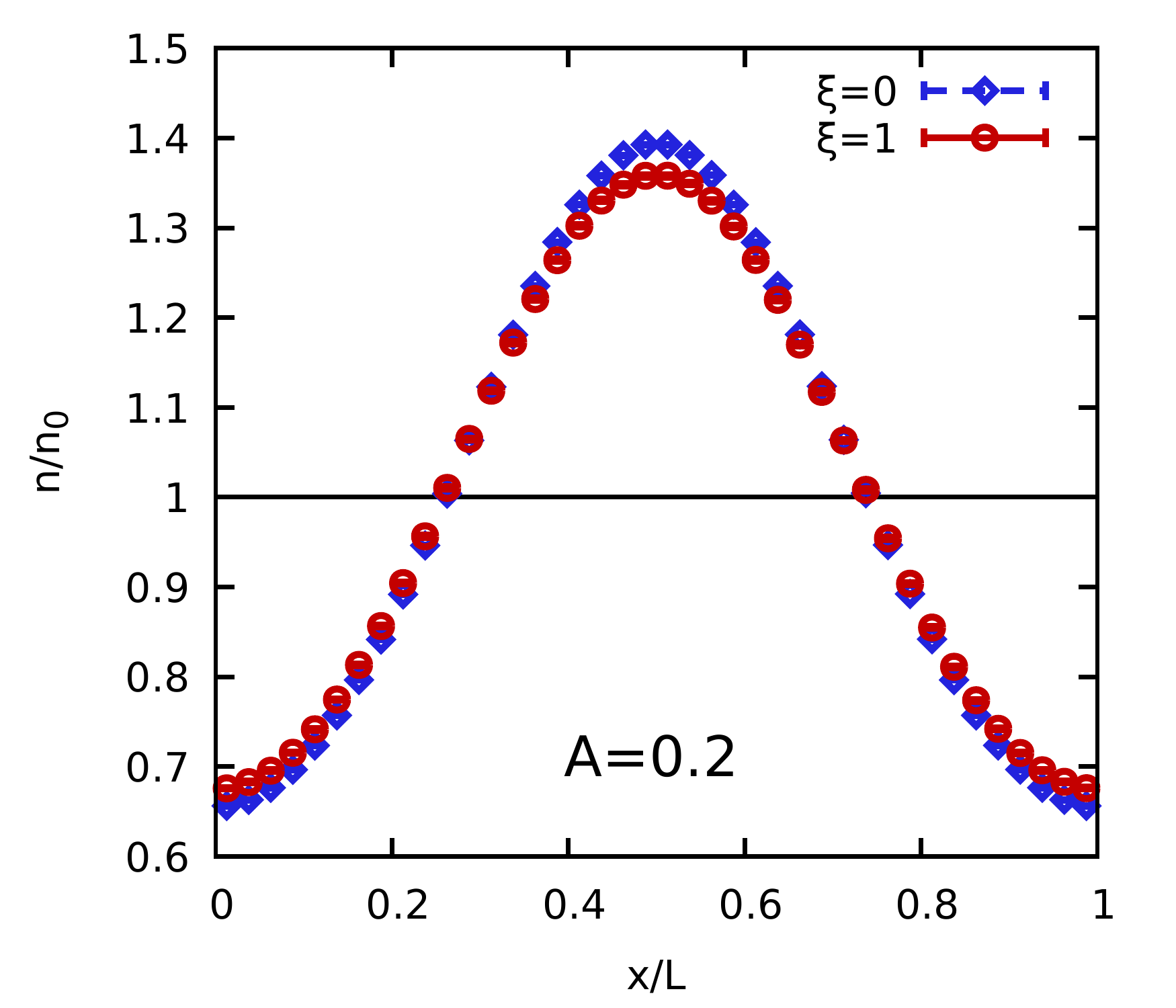}\hspace*{-0.03\textwidth}\includegraphics[width=0.51\textwidth]{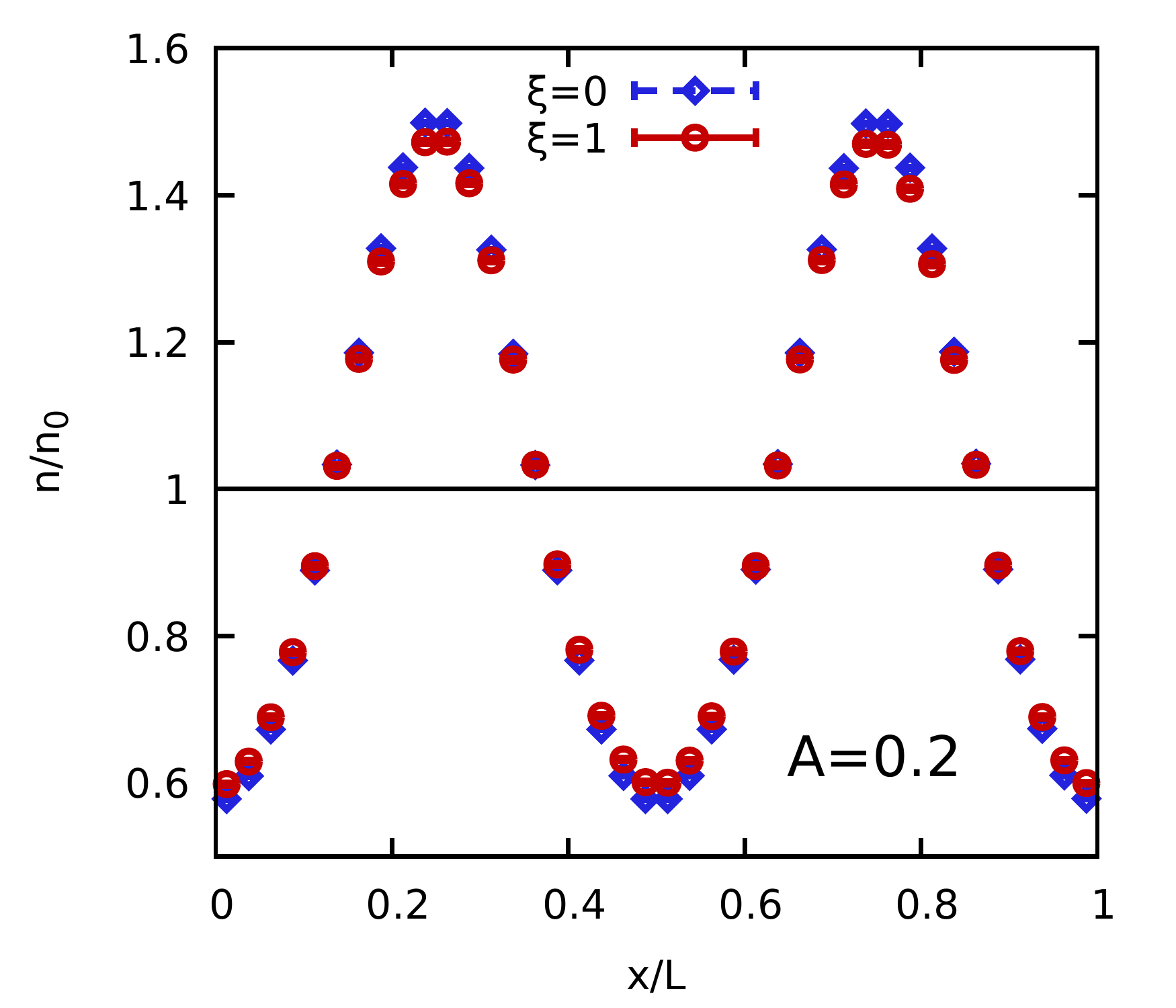}\\\vspace*{-1cm}
     \includegraphics[width=0.48\textwidth]{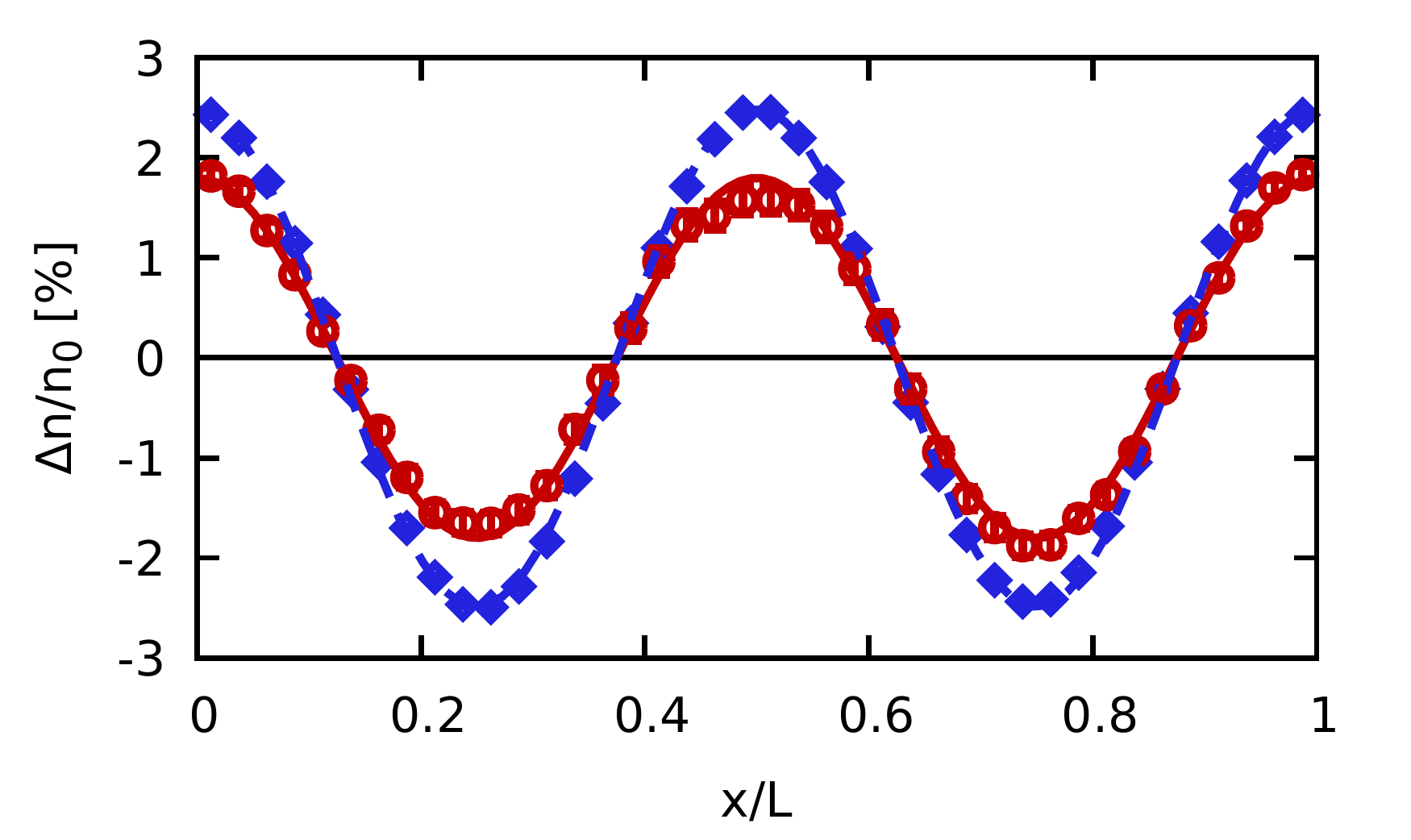}\includegraphics[width=0.48\textwidth]{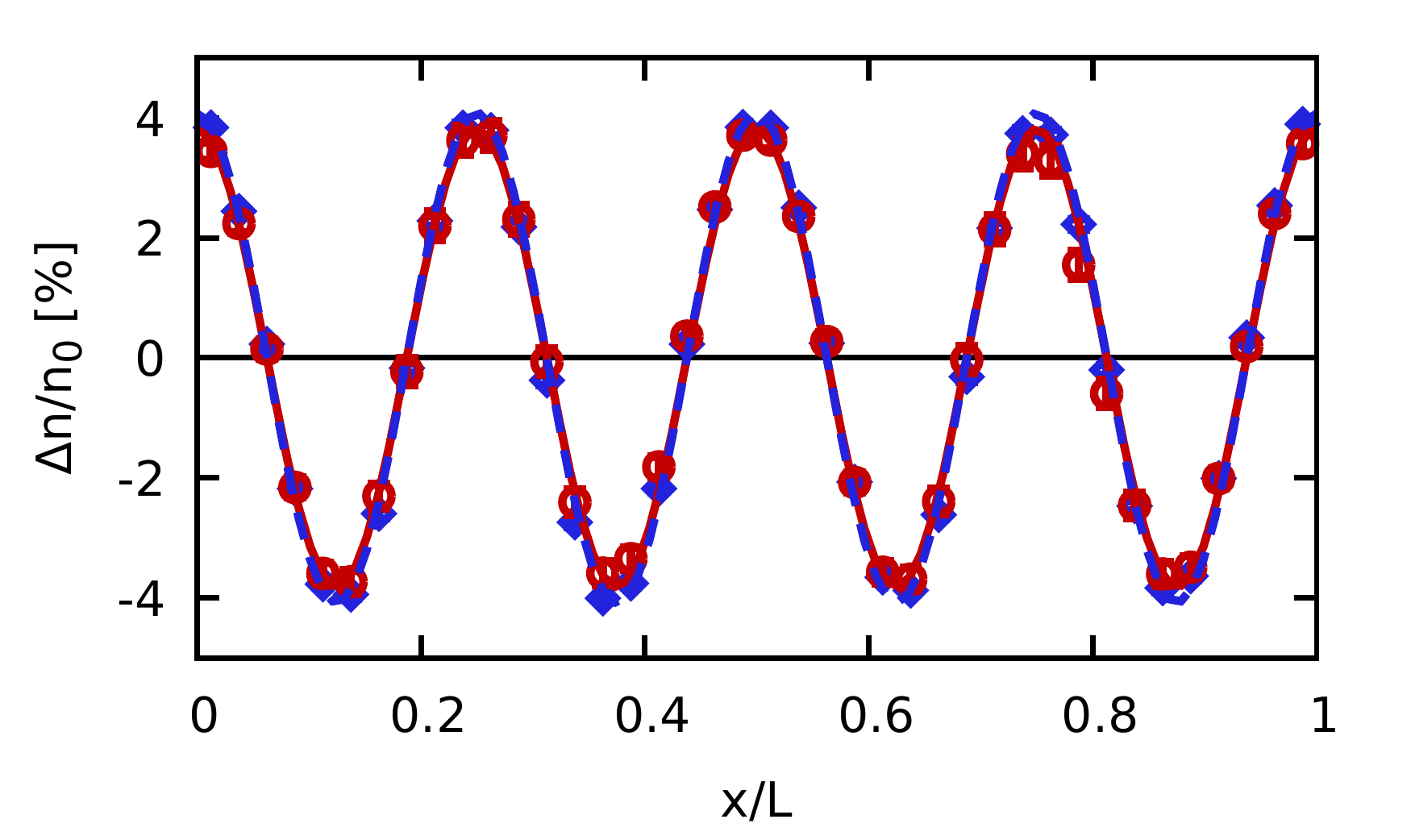}
    \caption{\label{fig:density_qx1}Density profile of the UEG for $N=14$, $r_s=2$, and $\theta=1$ with $q=2\pi/L$. The red and blue symbols depict our new data for $\xi=1$ and data for $\xi=0$ taken from Ref.~\cite{Dornheim_PRR_2021}.
    }
\end{figure}

Let us next investigate the corresponding density in coordinate space, which is depicted in Fig.~\ref{fig:density_qx1} for the perturbation amplitude $A=0.2$. The left (right) panel again corresponds to $q=0.84q_\textnormal{F}$ ($q=1.69q_\textnormal{F}$), and we show the density profile along the direction of the perturbation. Moreover, the red circles and blue diamonds correspond to $\xi=1$ and $\xi=0$.
In first order, the electrons do follow the external perturbation and cluster at those positions where the potential is at a minimum. Furthermore, the changes in the density compared to the unperturbed value $n_0$ exceed $30\%$ for all depicted cases, which is a strong indication that LRT is no longer accurate. This can be seen particularly well in the bottom row of Fig.~\ref{fig:density_qx1}, where we show the relative deviation between the PIMC data and the expansion in Eq.~(\ref{eq:expansion}) truncated after $\eta=1$. As nonlinear effects on the first harmonic are still small for the selected value of $A$ (see also Figs.~\ref{fig:A_dependence_qx1} and \ref{fig:A_dependence_qx2} below), the deviations are mainly due to the excitation at the second harmonic. Indeed, the depicted $\Delta n/n_0$
curves oscillate twice as fast as the original perturbations. 
Lastly, we again find that the impact of the spin-polarization is more pronounced for the smaller wave vector.

\begin{figure}
    \centering
 \includegraphics[width=0.48\textwidth]{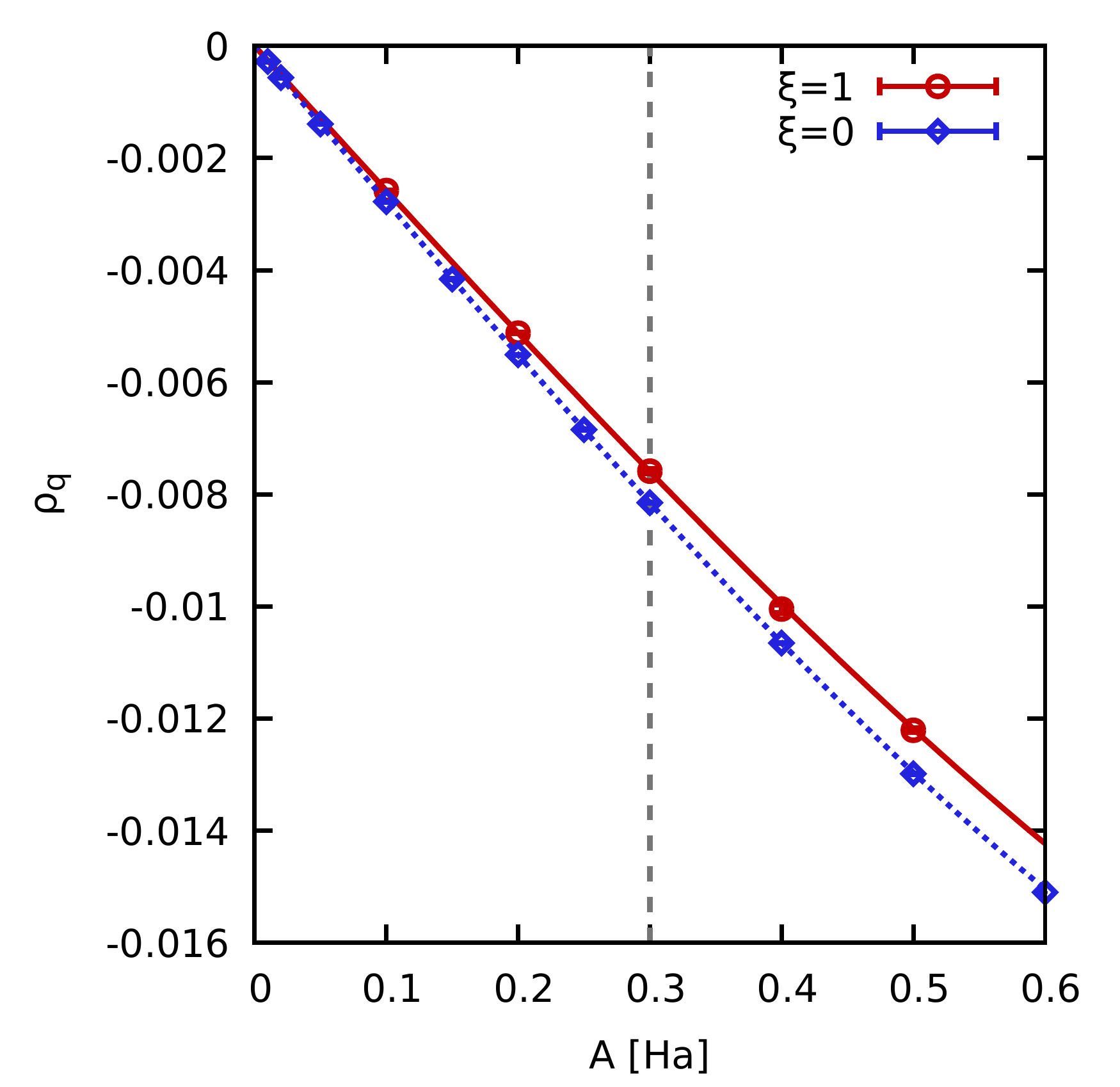}\includegraphics[width=0.48\textwidth]{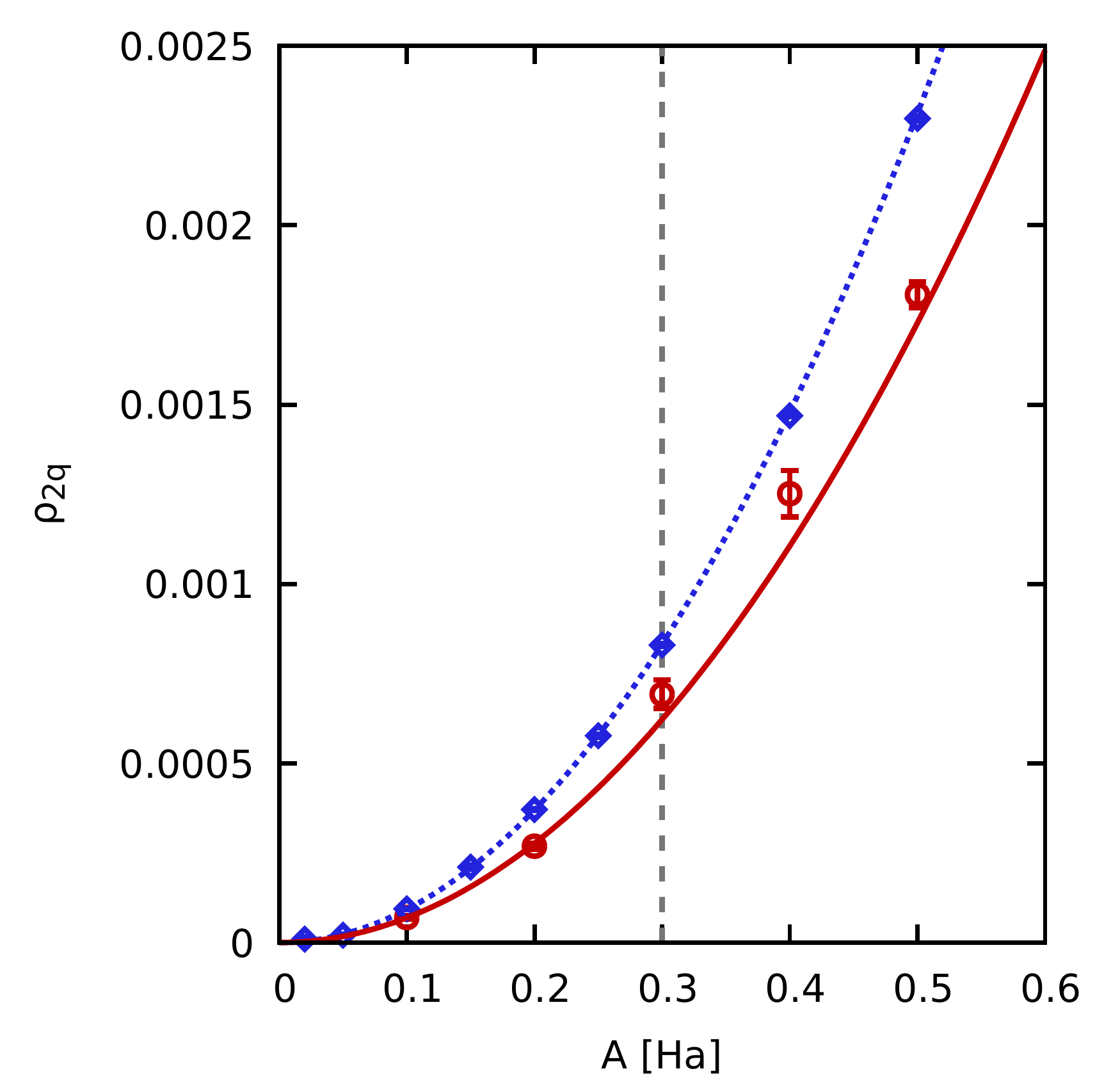}
    \caption{\label{fig:A_dependence_qx1} $A$-dependence of the UEG for $N=14$, $r_s=2$, and $\theta=1$ with $q=2\pi/L$. The red and blue symbols depict our new data for $\xi=1$ and data for $\xi=0$ taken from Ref.~\cite{Dornheim_PRR_2021}.
    }
\end{figure}

A more systematic investigation of the dependence of the nonlinear density response on the perturbation amplitude $A$ is shown in Fig.~\ref{fig:A_dependence_qx1} for the smaller wave number, $q=0.84q_\textnormal{F}$. More specifically, the left panel shows the density response [Eq.~(\ref{eq:rho})] at the first harmonic, $\mathbf{k}=\mathbf{q}$ with the red circles and blue diamonds showing the actual PIMC data for $\xi=1$ and $\xi=0$. Further, the corresponding curves show fits to these data according to Eq.~(\ref{eq:rho1}) in the interval $A\in[0,0.3]$ (see the vertical dashed grey line), with $\chi^{(1)}$ and $\chi^{(1,\textnormal{cubic})}$ being the free parameters. 
First and foremost, we note that the fitted curves are in excellent agreement to the PIMC data points well beyond the fitting range, which is a strong empirical confirmation of the functional form given in Eq.~(\ref{eq:rho1}). Furthermore, the density response of the ferromagnetic case is systematically smaller compared to the response of the paramagnetic system, and the deviation increases with $A$.

The same effect can be seen in the right panel of Fig.~\ref{fig:A_dependence_qx1}, where we show the signal at the second harmonic, $\mathbf{k}=2\mathbf{q}$. In this case, the curves have been obtained by fitting Eq.~(\ref{eq:rho2}) to the PIMC data, with the quadratic density response function $\chi^{(2)}$ being the single free parameter. Again, the fit nicely reproduces the input data, as the actual response at the second harmonic is indeed parabolic in leading order. Moreover, we note that the effect of the spin-polarization is even more pronounced in this case compared to the first harmonic.

\begin{figure}
    \centering
 \includegraphics[width=0.48\textwidth]{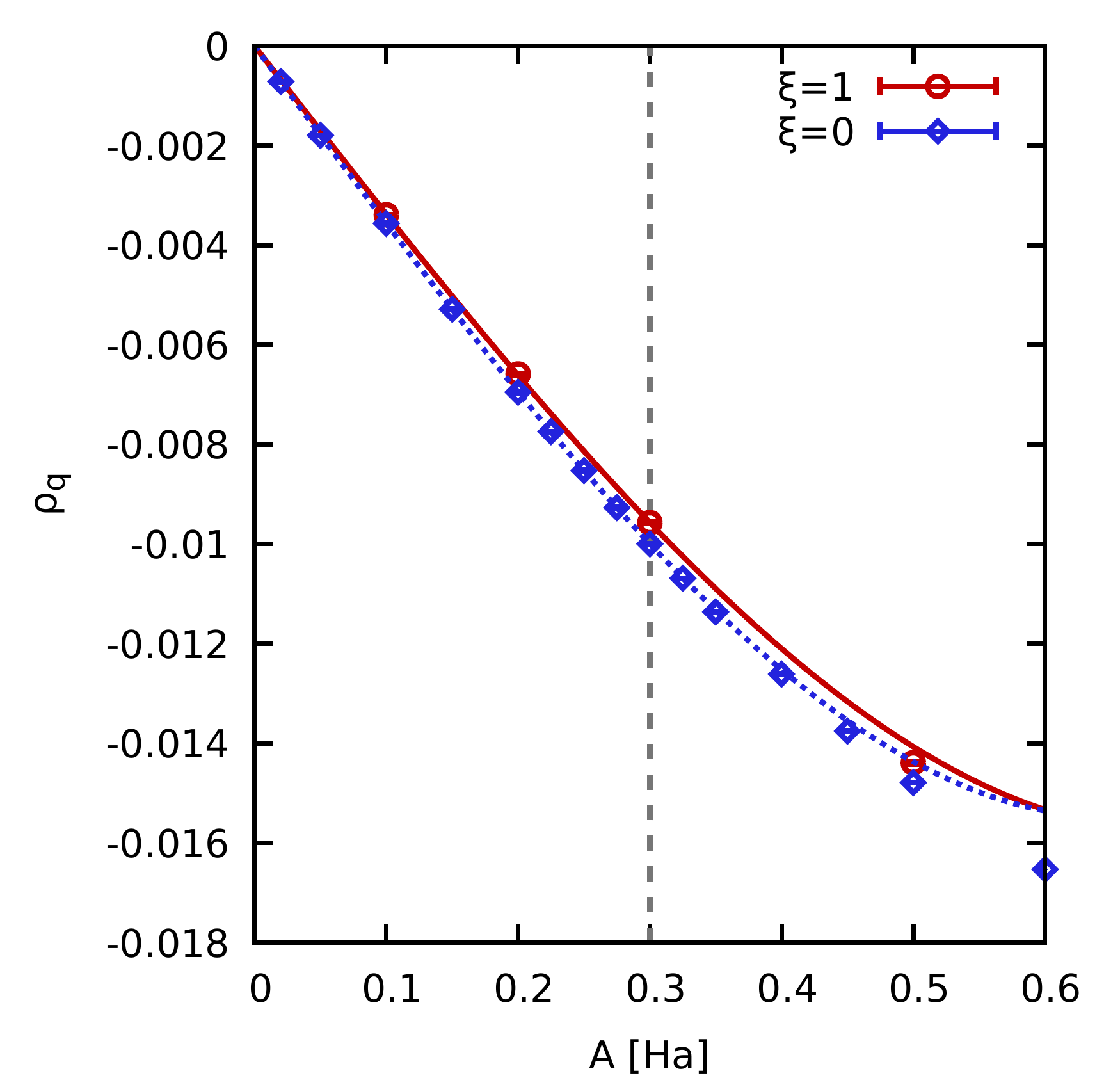}\includegraphics[width=0.48\textwidth]{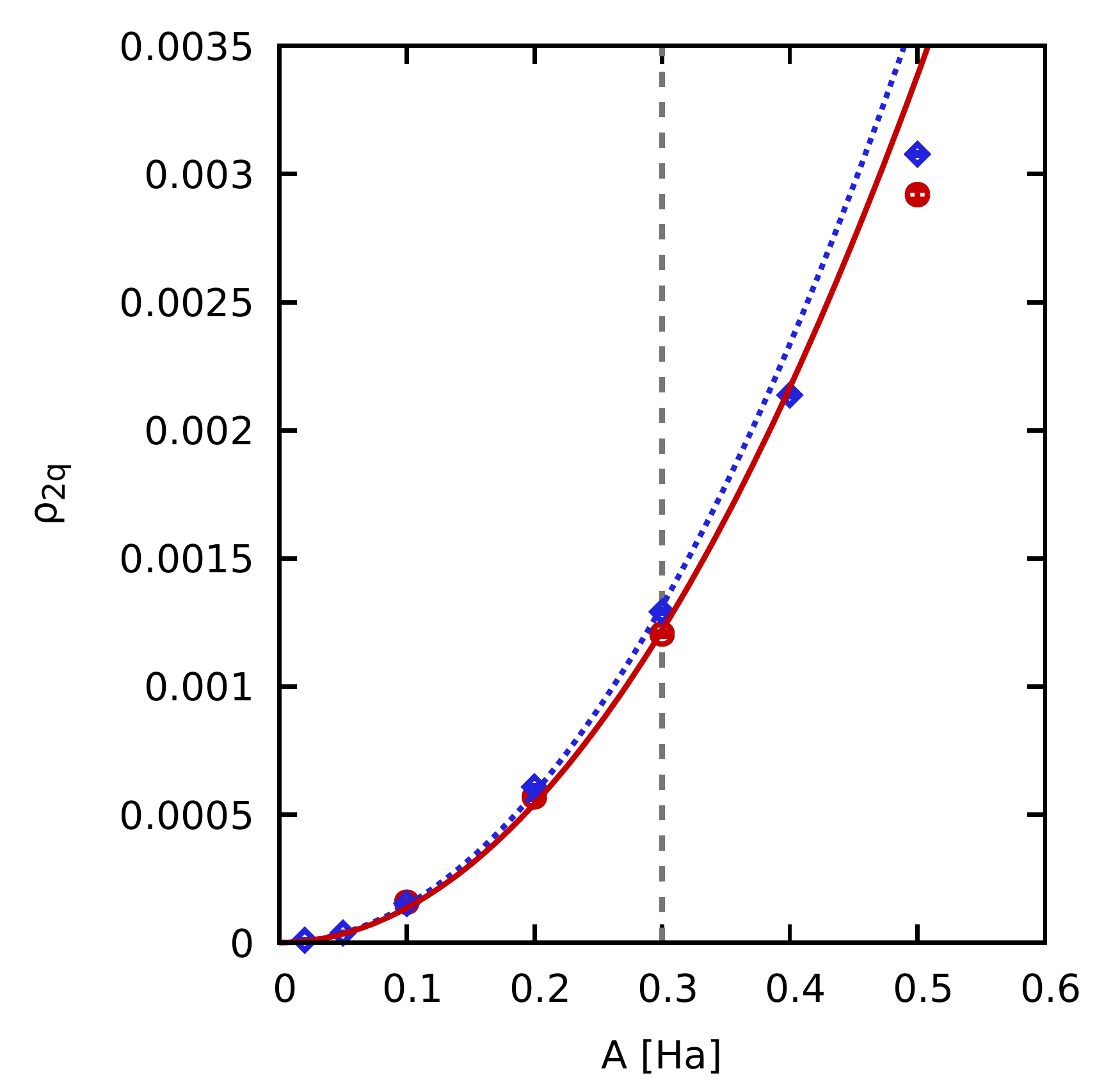}
    \caption{\label{fig:A_dependence_qx2} $A$-dependence of the UEG for $N=14$, $r_s=2$, and $\theta=1$ with $q=2\pi/L$. The red and blue symbols depict our new data for $\xi=1$ and data for $\xi=0$ taken from Ref.~\cite{Dornheim_PRR_2021}.
    }
\end{figure}

As a next step, we proceed with an analysis of the same information for the larger wave number ($q=1.69q_\textnormal{F}$), which is depicted in Fig.~\ref{fig:A_dependence_qx2}. In this case, the $A$-dependence of the first harmonic more strongly deviates from a straight line compared to Fig.~\ref{fig:A_dependence_qx1}, as the cubic response function $\chi^{(1)}$ is substantially larger in magnitude~\cite{Dornheim_PRR_2021}. Still, the functional form given in Eq.~(\ref{eq:rho1}) remains accurate. At the same time, the impact of the spin-polarization at the first harmonic is small both in the linear and nonlinear regime. 
The $A$-dependence of the density response at the second harmonic shown in the right panel of Fig.~\ref{fig:A_dependence_qx2} exhibits a similar behaviour, although the fits become inaccurate for smaller values of the perturbation amplitude (compared to Fig.~\ref{fig:A_dependence_qx1}), where terms of a higher order in $A$ that have been neglected in Eq.~(\ref{eq:rho2}) start to become important.

Let us conclude this investigation of the impact of the spin-polarization on the nonlinear density response of the UEG with a discussion of the physical origin of the comparably smaller spin-dependence at $q=1.69q_\textnormal{F}$ compared to $q=0.84q_\textnormal{F}$. 
A possible explanation for this finding are electronic exchange--correlation effects, which are well-known to be most pronounced around $q=2q_\textnormal{F}$. As correlations tend to suppress spin-effects~\cite{dornheim2021momentum}, this would explain the small manifestation of the latter on the nonlinear density response for the larger wave number.
A second possible explanation for this empirical finding is given by the characteristic length scales of the system. More specifically, large wave numbers $q$ correspond to small distances in coordinate space, which means that the physical properties of the UEG at large $q$ are mainly determined by single-particle effects. Since spin-effects, by definition, must involve multiple particles, this would also explain the smaller impact of $\xi$ for $q=1.69q_\textnormal{F}$.

\section{Summary and Outlook\label{sec:summary}}

In this work, we have presented the first \emph{ab initio} PIMC results for the nonlinear density response of the ferromagnetic electron gas at WDM conditions. The comparison to previous results for the paramagnetic case has revealed that spin-effects cannot be neglected at parameters that are of relevance for contemporary experiments. More specifically, the stronger degree of correlations between identical fermions for $\xi=1$ leads to a reduced density response both in the linear and in the nonlinear regime. Moreover, the particular impact of the spin-polarization strongly depends on the wave vector $\mathbf{q}$ in a nontrivial way. 

Unfortunately, the computation of the nonlinear response functions over the entire relevant range of wave numbers is computationally unfeasible due to the substantially more severe manifestation of the fermion sign problem in the ferromagnetic case. This problem might be overcome in the future by estimating and subsequently integrating a new class of generalized imaginary-time correlation functions, which give access to the full nonlinear density response from a single simulation of the unperturbed system~\cite{dornheim2021nonlinear}. An additional topic for future research is given by the consideration of intermediate values of the spin-polarization of $\xi\in[0,1]$, which might be accompanied by the investigation of the spin-resolved nonlinear density response of the individual components.

\section*{Data Availability Statement}
The data that support the findings of this study are available from the corresponding author upon reasonable request.

\section*{Acknowledgements}
This work was funded by the Center for Advanced Systems Understanding (CASUS) which is financed by Germany's Federal Ministry of Education and Research (BMBF) and by the Saxon Ministry for Science, Culture and Tourism (SMWK) with tax funds on the basis of the budget approved by the Saxon State Parliament.
We gratefully acknowledge CPU-time at the Norddeutscher Verbund f\"ur Hoch- und H\"ochstleistungsrechnen (HLRN) under grant shp00026 and on a Bull Cluster at the Center for Information Services and High Performance Computing (ZIH) at Technische Universit\"at Dresden.


\bibliography{ref}

\end{document}